\documentclass[%
 reprint,
superscriptaddress,
showpacs,preprintnumbers,
 amsmath,amssymb,
]{revtex4-1}

\usepackage{graphicx}
\usepackage{dcolumn}
\usepackage{bm}

\usepackage{color,soul}
\begin{document}


\title{Robust entanglement of an asymmetric quantum dot molecular system in a Josephson junction }%

\author{E. Afsaneh}%
 \affiliation{Department of Physics, Faculty of Science, University of Isfahan, Hezar Jerib Str., Isfahan 81746-73441, Iran.}
\author{M. Bagheri Harouni}%
\affiliation{Department of Physics, Faculty of Science, University of Isfahan, Hezar Jerib Str., Isfahan 81746-73441, Iran.}%
\affiliation{Quantum Optics Group, Department of Physics, Faculty of Science, University of Isfahan, Hezar Jerib Str., Isfahan 81746-73441, Iran}%

\date{\today}

\begin{abstract}
We demonstrate how robust entanglement of quantum dot molecular system in
a voltage-controlled junction can be generated. To improve the
quantum information characteristics of this system, we propose an
applicable protocol which contains the implementation of asymmetric
quantum dots  as well as engineering reservoirs. Quantum dots with
tunable energy barriers can provide asymmetric coupling coefficients
which can be tuned by gap voltages. Also by engineering reservoirs,
superconductors can be used as leads in a biased-voltage junction.
The high-controllability characteristics of system supplies the
arbitrary entanglement by tuning the controlling parameters.
Significantly in concurrence-voltage characteristics, perfect entanglement can be achieved in an asymmetric
structure and it can be kept with near-unit magnitude in response to bias voltage increasing.
\end{abstract}

\maketitle


\section{Introduction}
Recent advancements in condensed matter physics and
nanotechnology open new possibilities for the implementation of
nanodevices in quantum information studies. Although, the concept of
entanglement was basically studied for distinguishable bipartite
systems\cite{Peres,EntanglementDistillation, Werner,Horodecki,
Nielsen}, in recent years there has been a great deal of
interest in quantifying the entanglement of indistinguishable
components in condensed matter systems. The elements of these
systems are identical massive particles which involve quantum
correlations at short distances. The entanglement of
indistinguishable particles either bosons or fermions should be
characterized with their symmetrized or antisymmetrized wave
functions
respectively\cite{Schliemann-2001,Schliemann-2001-2,Schliemann-2002,Zanardi-2001,Zanardi-2002,Wiseman-2003,Wiseman-2006,Fazio,Majtey,Rossignoli-2018,Compagno}.

Particularly, the entanglement of fermions in condensed matter
systems can be evaluated by two methods: entanglement of
modes\cite{Zanardi-2001,Zanardi-2002,Rossignoli-2018,Enk,Benatti,Puspus}
and entanglement of
particles\cite{Schliemann-2001,Schliemann-2001-2,Schliemann-2002,Wiseman-2003,Wiseman-2006,Majtey}.
In the former, the entanglement of indistinguishable fermions is
associated with the shared modes not particles of subsystems in
single particle Hilbert space.  But for the latter, the entanglement
of fermions specifically is concerned about the antisymmetrization
of quantum wave functions of indistinguishable fermionic particles.

For fermionic entanglement of particles approach, firstly the
quantum correlations of two fermions in a 2K dimensional
single-particle space were characterized\cite{Schliemann-2001}.
After that more than two indistinguishable particles fermions in
higher-dimensional single-particle spaces were analyzed and quantum
correlations of pure states in the arbitrary-dimensional Hilbert
space were classified\cite{Schliemann-2002}. Recently, a
multipartite concurrence was introduced for N-indistinguishable
fermionic particles in an arbitrary-dimensional pure
states\cite{Majtey}. It was presented that the multipartite
concurrence can be displayed as an average amount of one observable
when two copies of the compound state are accessible.

For studying the entanglement of indistinguishable fermionic particles, quantum dots(QDs)\cite{Loss,Hanson-2} can be taken into account as promising candidates. QDs as one branch of broad two-state qubit systems\cite{Nori-2} play prominent roles in nanostructures for their tunable discrete energy levels and also for their easy controllability of barriers by gate voltages.

Also, quantum dot molecules (QDMs) consist of quantum dots which are coupled by tunneling and separated by barriers have received great attention theoretically and experimentally\cite{Borges,Carlson,Bayer,Temirov}. These quantum structures have been selected as the ideal choices for researching the quantum information processing.
The analysis of entanglement dynamics between two electrons inside coupled quantum molecules demonstrated the crucial entanglement characteristics\cite{Oliveiraa}.

Theoretical \cite{Daroca} and experimental \cite{Brackera} studies
showed that asymmetric structure of quantum molecules has enhanced the
control of tunneling features. It was theoretically shown that in an asymmetric
quantum dot molecular system, the fidelity of entangled photon pairs can be achieved near-unit magnitude\cite{Jennings}. In
addition, the asymmetric quantum dot-lead couplings have been
extensively implemented in electrical \cite{Malz} and thermal
\cite{Tang} rectification devices to improve the electric and heat
transport technologies.

Moreover, superconducting devices have found impressive interest in
quantum information setups \cite{Song,Dickel,Wang} because of their
long intrinsic coherency with no dissipation characteristics. Recent
years, employing the superconducting qubits and superconducting
resonators have improved the exploring of entanglement \cite{Nori,Egger},
teleportation \cite{Roos,Riebe,Barrett} and quantum computing
\cite{Majer,Houck,Schuster} studies. Superconducting qubits namely
phase \cite{Martinis}, flux \cite{Friedman,van derWal} and charge
\cite{Yang,Nakamura} qubits can be connected with microwave
\cite{Johansson}, electrical \cite{Stern,Ma}, mechanical
\cite{Connell} and superconducting \cite{Reuther} resonators.
According to the frequency range of superconducting devices, these
nanostructures would be driven by microwave
\cite{Gambetta,Blais-2007} or optical \cite{Santos,Wol} fields. Also, QDs
in normal biased-voltage junctions have extensively been used
experimentally \cite{Reed,Park,Dadosh} and theoretically
\cite{Stafford,Nitzan,Smit}.
Recently, quantum transport through the QDs system in contact with Josephson junctions (JJs) which act as the single transistors to filter the transfer of electrons have attracted a great deal of attention\cite{Yeyati, Konig, Kosov,Grifoni,Afsaneh}.

It seems that quantum information studies on an asymmetric quantum dot molecule in a bias-controlled Josephson junction can be considered as an interesting area for research which can provide novel achievements. Therefore in this study, we propose a QDM system in a conventional JJ with asymmetric tunneling coefficients to achieve the robust entanglement
and also to keep its magnitude near-unit under the bias voltage
control. To this end, we consider the indistinguishable
entanglement for our system which becomes possible by evaluating the
fermionic concurrence. To explore the quantum information processing of QDM system in a
biased-voltage junction, we perform our analysis in Markovian
regime. First, we obtain the quantum transport of molecular system
to show the current-voltage characteristics (I-V) as one of the
important properties of biased-voltage circuits. Then, we
investigate the control of the entanglement with respect to  bias
voltage. We find that with only bias voltage control, the complete
controllability to yield perfect entanglement is not possible.
Therefore, we apply the strategy of left-right asymmetric coupling
strength to achieve the robust entanglement. The dynamics of
entanglement and its response to bias voltage in different
situations of symmetric and asymmetric couplings demonstrate a wide
flexibility of the proposed setup to provide a desired high
entanglement. The main advantage of this molecular system includes
the feasible controlling elements of easy-tunable bias voltage
driving field and the manipulation of quantum dot couplings. Indeed
by engineering reservoirs and the presence of superconducting leads,
the performance of system is extensively influenced to provide
robustly entangled states.

This paper is organized as follows: In Sec.(\ref{Model}), we
introduce the proposed model composed of a quantum dot molecular
system in a JJ by describing the whole Hamiltonian. We compute the
quantum transport of our molecular system in Sec.(\ref{Current}). In
Sec.(\ref{Concurrence}) by introducing symmetric and asymmetric
structures, we obtain the entanglement of QDM system under the bias
voltage control. In Sec(\ref{Results}), we present the results of
the entanglement behavior in bias voltage changes and its time
evolution in constant bias voltages and also for specific order
parameters.  Finally, we conclude the results in
Sec.(\ref{Conclusion}). In Appendix A, we describe how to
diagonalize the Hamiltonian of superconducting leads by applying the
Bogoliubov transformation. In Appendix B, we calculate the quantum
master equation to study the dynamics of system.
\section{Model} \label{Model}
\begin{figure}[t]
\centering
\includegraphics[scale=0.15]{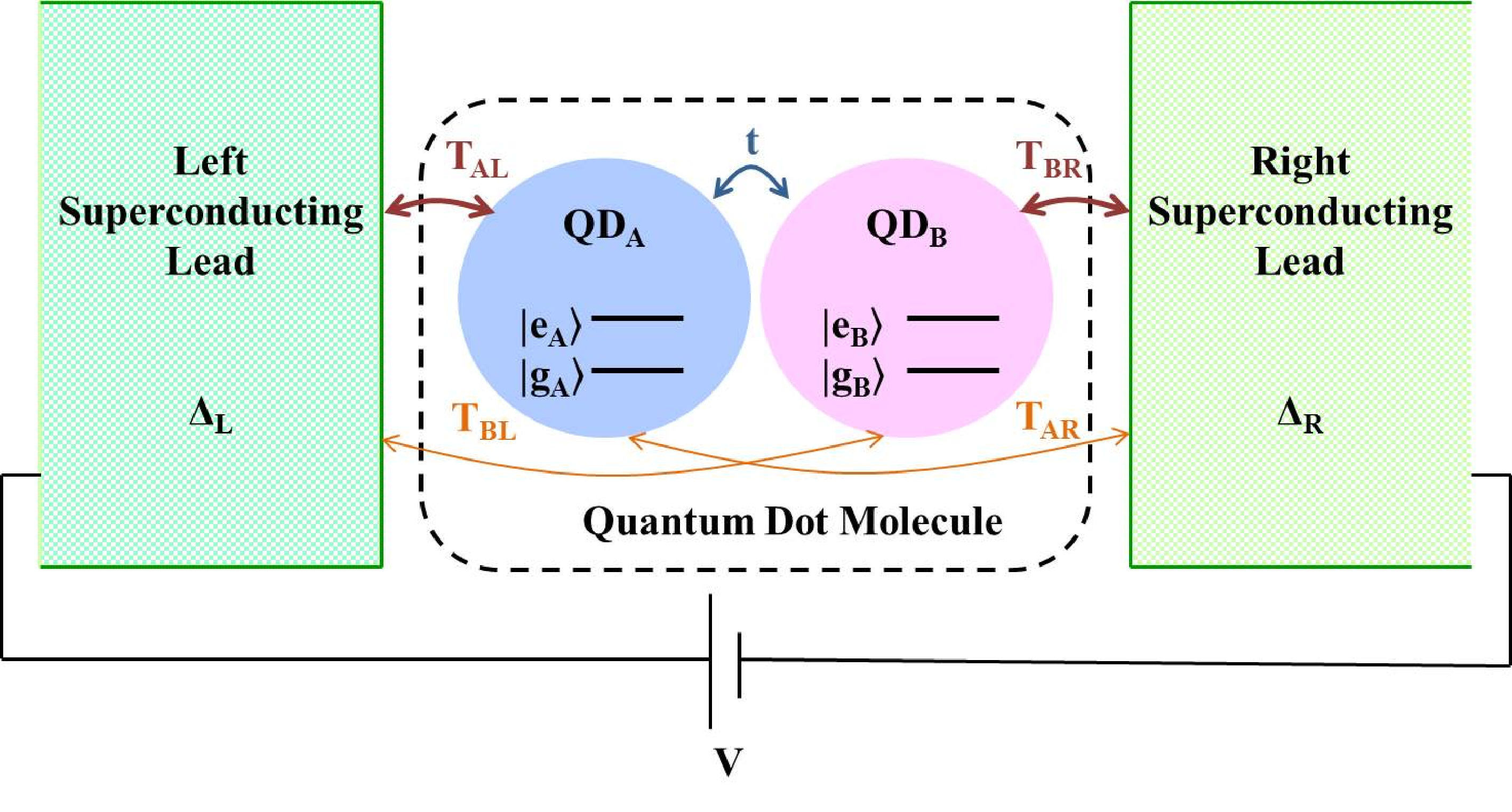}
\caption{The proposed physical system: A quantum dot molecule system
consists of two coupled quantum dots, $A$ and $B$, with inter-dot
coupling strength $t$ and QD-lead coupling strengths: $T_{AL}$, $T_{AR}$, $T_{BR}$, $T_{BL}$. The superconducting leads
are under the bias voltage $V$.}\label{Fig1}
\end{figure}

The proposed open quantum system consists of a QDM weakly coupled to
the superconducting leads which is demonstrated in Fig.(\ref{Fig1}),
schematically. Applying an external bias voltage between the leads
$L$ and $R$ induces the electron transport from the left to the
right.
The Hamiltonian of the whole system can be written as:
\begin{equation}\label{eq1}
\hat{H}=\hat{H}_{QDM}+\hat{H}_{Leads}+\hat{H}_{int},
\end{equation}
For simplicity, the molecular quantum dot system is taken in spinless Anderson-Holstein model\cite{Jovchev,Khedri}. So, $\hat{H}_{QDM}$ is expressed as:
\begin{equation}
\hat{H}_{QDM}=\sum_{\alpha} \varepsilon_{\alpha}\hat{d}^{\dagger}_{\alpha}\hat{d}_{\alpha}
+t(\hat{d}^{\dagger}_{A}\hat{d}_{B}+\hat{d}_{A}\hat{d}^{\dagger}_{B}).
\end{equation}
here, $\hat{H}_{QDM}$ introduces the spinless double quantum dot(DQD) with electronic energy levels $\varepsilon_{\alpha}$ for $\alpha=A, B$. Beyond the Coulomb blockade regime, each QD is considered in single electron conditions\cite{Averin,Beenakker}. In the second term, $t$ describes the inter-dot hopping strength which can be tuned using an applied gate voltage.
In Eq.(\ref{eq1}),
$\hat{H}_{Leads}$ corresponds to the left and right superconducting
leads which are described by the mean-field Hamiltonian as
\cite{deGenne, Tinkham}:
\begin{equation}\label{MFH}
\hat{H}^{MF}_{Leads}=\sum_{k \nu \sigma}  \xi_{k \nu}  \hat{c}^{\dagger}_{k \nu \sigma} \hat{c}_{k \nu \sigma}+\sum_{k \nu} \Big( \Delta_{\nu} \hat{c}^{\dagger}_{k \nu \uparrow} \hat{c}^{\dagger}_{-k \nu \downarrow}+ \Delta^{*}_{\nu} \hat{c}_{-k \nu \downarrow} \hat{c}_{k \nu \uparrow} \Big).
\end{equation}
Here, $\hat{c}^{\dagger}_{k \nu \sigma} ( \hat{c}_{k \nu \sigma} ) $
is the creation (annihilation) operator of an electron with momentum
$k$ and spin $\sigma={\uparrow, \downarrow}$ in lead $\nu=L,R$. In
this relation, $\xi_{k \nu}=\varepsilon_k - \mu_{\nu}$ is the
particle energy in which $\varepsilon_k$ denotes the single-particle
energy regards to the electrochemical potential $\mu_\nu$.
Moreover, $\Delta_{\nu}=|\Delta_{\nu}|e^{i \phi_{\nu}}$ remarks the
superconducting energy gap of lead $\nu$ with the superconducting
phase, $\phi_\nu$.
The mean field Hamiltonian could be diagonalized by applying
Bogoliubov transformation to obtain (Appendix A):
\begin{equation}
\hat{H}_{Leads}=E_G+\sum_{k \nu \sigma} E_{\nu k}  \hat{\gamma}^{\dagger}_{k \nu \sigma} \hat{\gamma}_{k \nu \sigma},
\end{equation}
where $E_G$, the ground state energy, represents the Cooper pair condensate energy.
The interaction Hamiltonian, $\hat{H}_{int}$ in Eq.(\ref{eq1}),
corresponds to the tunneling between the QDs and electrodes which
can be written as:
\begin{equation}
\hat{H}_{int}=\sum_{k \nu \alpha} \left(T_{k \nu \alpha} \hat{c}^{\dagger}_{k \nu } \hat{d}_{\alpha} +T^{*}_{k \nu \alpha} \hat{c}_{k \nu}
\hat{d}^{\dagger}_{\alpha}\right).
\end{equation}
The tunneling coefficient, $T_{k \nu \alpha}$, describes
the coupling strength depending on $k$, the momentum of an electron
in lead $\nu$, the site of quantum dot $\alpha$.

To investigate  the time evolution of system, firstly the quantum master equation(QME)
and the density matrix are obtained (Appendix B) and then we
calculate the current and entanglement in the following sections.
\section{Current}\label{Current}
\begin{figure}[t]
\centering
\includegraphics[scale=0.2]{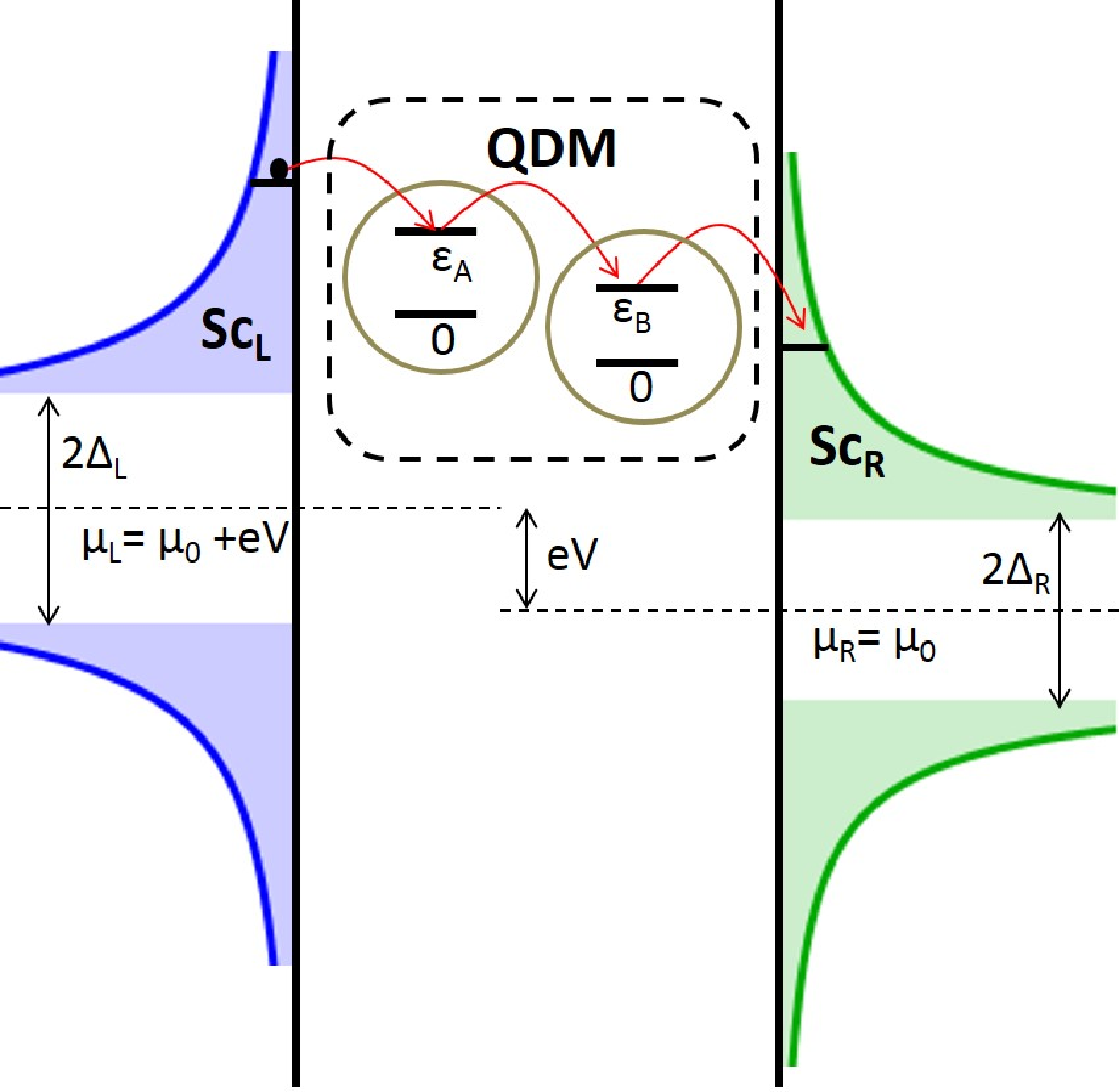}
\caption{The density of states in the superconducting reservoirs of $Sc_L/QDM/Sc_R$ junction. The asymmetric applied bias voltage lets carriers to flow from the left reservoir to the QDM and then to the right lead.}\label{Fig2}
\end{figure}

In order to have transport through the present system(Fig. (\ref {Fig1})), the asymmetric bias voltage is applied to the electrodes which is shown in Fig. (\ref {Fig2}). External bias voltage changes the density of states such that the electrochemical energy level of the left lead is shifted higher than the energy levels of the QDM and the right lead which causes flowing current through the junction.   

\begin{figure}[t]
\centering
\includegraphics[scale=0.8]{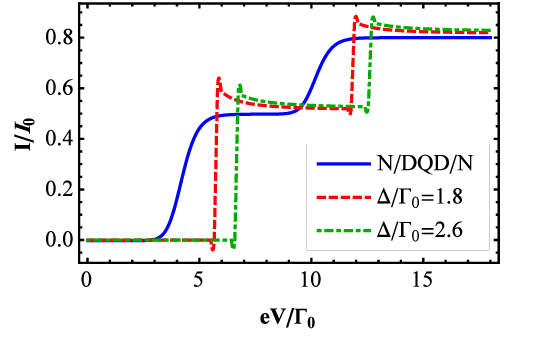}
\caption{Current-voltage characteristics in specific
superconducting energy gaps. Normal leads: Solid line $\Delta=0$,
Superconducting leads: Dashed line $ \frac{\Delta}{\Gamma_0} =1.8$,
Dotdashed $\frac{\Delta}{\Gamma_0}=2.6$ when $\Gamma_0=\pi N_F
|T|^2$ , $I_0=e\frac{\Gamma_0}{\hbar}$ and $\Delta_L=\Delta_R=\Delta$.  }\label{Fig3}
\end{figure}

Current as a measurable quantity denotes the variation of particle
number, $N$, in lead $\nu$ which is defined as\cite{Mahan}:
\begin{eqnarray} \label{current}
\hat{I}_{\nu}(t)&=& -e \frac{d \hat{N}_{\nu}}{dt}=\frac{i e}{\hbar} [\hat{N}_{\nu}(t), \hat{H}_{I}(t)]\nonumber \\
&=& \frac{ie}{\hbar} \sum_{k \alpha} (T_{k \alpha} \hat{c}^{\dagger}_{k \nu} \hat{d}_{\alpha} - T^{*}_{k \alpha} \hat{c}_{k \nu} \hat{d}^{\dagger}_{\alpha}),
\end{eqnarray}
where $\hat{N}_{\nu}=\sum_{\nu} \hat{c}^{\dagger}_{\nu}
\hat{c}_{\nu} $. According to the QME formalism,
Eq.(\ref{MasterEq}), the density matrix evolution of the system
would be written as $ \dot{\hat{\rho}}=\hat{M} \hat{\rho}$. In this
relation matrix $\hat{M} $ shows the properties of master equation.
Therefore, we can rewrite the current formula, Eq.(\ref{current}),
as \cite{Afsaneh, Mukamel}:
\begin{equation}
\hat{I}_{\nu}(t)= \frac{e}{\hbar} \langle \hat{N} | \hat{M}_{\nu} | \hat{\rho}(t) \rangle,
\end{equation}
where $\hat{M}_{\nu} $ shows the contribution of lead $\nu$ in
matrix $\hat{M} $. In steady state of the system, by taking so long
time ($t \rightarrow \infty $), the stationary transport is shown in
Fig.(\ref{Fig3}) containing the plots of normal junction
($\Delta=0$) and JJ with different energy gaps.

According to I-V characteristics which is shown in Fig.(\ref{Fig3}), the magnitude of current is growing
by the increase of bias voltage. Only in energies equal to the
quantum dots' energ levels, the current hits the peaks in delta type for
the superconducting leads while it illustrates the smooth steps for
the normal leads. Although, the current level of system is increased
by rising the magnitude of energy gaps, it reaches the platform for
the large enough bias voltage.

In all calculations in order to deal with only the quasiparticle transport and ignoring the Cooper pair current, we assume all energy levels are far enough from the order parameter of leads. 
\section{Concurrence}\label{Concurrence}
It is convenient to apply the concurrence as a measure of
entanglement for two qubit systems. In the following, first this
measure of entanglement for two distinguishable qubits is defined.
Then, fermionic concurrence for indistinguishable particles
will characterized and evaluated in analogue with Wootters'
formula.
\subsection{Concurrence of distiguishable particles}
For the first time, Wootters introduced the measure of
concurrence to evaluate the entanglement of qubits with two parties
in both pure and mixed states
\cite{concurrence-1,concurrence-2}. This measure of entanglement
is defined as:
\begin{equation}\label{concurrence}
C(\rho)=Max [0,\lambda_1-\lambda_2-\lambda_3-\lambda_4],
\end{equation}
in which, $\lambda_i,(i=1,2,3,4)$ represents the non-negative
eigenvalues of a matrix $\hat{R}$ in decreasing order
$\lambda_1>\lambda_2>\lambda_3>\lambda_4$. The matrix $\hat{R}$ is defined as:
\begin{equation}
\hat{R}=\sqrt{\sqrt{\hat{\rho}}\tilde{\hat{\rho}}\sqrt{\hat{\rho}}},
\end{equation}
where $ \hat{\rho}=\sum_ip_i|\psi_i\rangle\langle\psi_i|$
denotes the density matrix of system in which, $p_i$ is the
probability of each state of decompositions. Also 
$\tilde{\hat{\rho}}=(\hat{\sigma}_y \otimes \hat{\sigma}_y)
\hat{\rho}^*(\hat{\sigma}_y \otimes \hat{\sigma}_y)$. In this
relation, $\hat{\sigma_y}$ describes the $y$ element of
Pauli matrices and $\hat{\rho}^*$ represents the complex
conjugate of the density matrix.
\subsection{Concurrence of indistinguishable fermions}
In condensed matter systems, the entanglement of electrons
should be taken into account as indistinguishable particles. To
characterize the entanglement of indistinguishable fermions, the
simplest possible system with the lowest-dimensional situation is
defined for two fermions in four-dimensional single-particle Hilbert
space which is generally in a six-dimensional two-particle
space\cite{Schliemann-2002}. An arbitrary state of two fermions
is given:
\begin{equation}
|\psi\rangle=\sum_{i,j=1}^{4}\psi_{i,j}\hat{c}^{\dagger}_{i}\hat{c}^{\dagger}_{j}|0\rangle
\end{equation}
where $\psi_{ij}$ indicates the coefficient matrix. Its
dual matrix
$\tilde{\psi}_{ij}=\frac{1}{2}\sum_{k,l=1}^{4}\varepsilon^{i,j,k,l}\psi^{*}_{k,l}$
is defined with antisymmetric unit tensor
$\varepsilon^{i,j,k,l}$. In this case, fermionic concurrence in
analogy with distinguishable two-qubit concurrence
Eq.(\ref{concurrence}) can be written
as\cite{Schliemann-2001,Schliemann-2001-2,Schliemann-2002,Fazio,
Majtey}:
\begin{eqnarray}
C_F(|\psi\rangle)&=&|\langle\tilde{\psi}|\psi\rangle|=\Big|\sum_{i,j,k,l=1}^{4}\varepsilon^{i,j,k,l}\psi_{i,j}\psi_{k,l}\Big|\nonumber\\
&=&8|\psi_{12}\psi_{34}+\psi_{13}\psi_{42}+\psi_{14}\psi_{23}|
\end{eqnarray}
Also, this relation can be expressed as\cite{Majtey,Rungta}:
\begin{equation}
C_F(|\psi\rangle)=\sqrt{2(1-2Tr[\hat{\rho}^2])}
\end{equation}
in which $\hat{\rho}$ denotes the single-fermion reduced density matrix.
This means that the Wootters’ formula Eq.(\ref{concurrence}) which states concurrence was proved completely for two indistinguishable fermions.     
\subsection{Concurrence in our system}
Due to the role of electrons in construction of QDs as qubits,
quantum dots are involved in fermionic statistics as well as
antisymmetric wave functions. Therefore to calculate the
entanglement of coupled QDs, it is needed to use the entanglement of
indistinguishable particles method. In our study, it is assumed that
quantum dots with spinless electrons can be realized as qubits by
their orbital electronic degrees of freedom in quantum information
theory. These two spinless electrons in a double-well potential are
close enough to each other in short distance to have quantum
correlations. Therefore, they can treat as indistinguishable
particles entanglement.

\begin{figure}[t]
\centering
\includegraphics[scale=0.2]{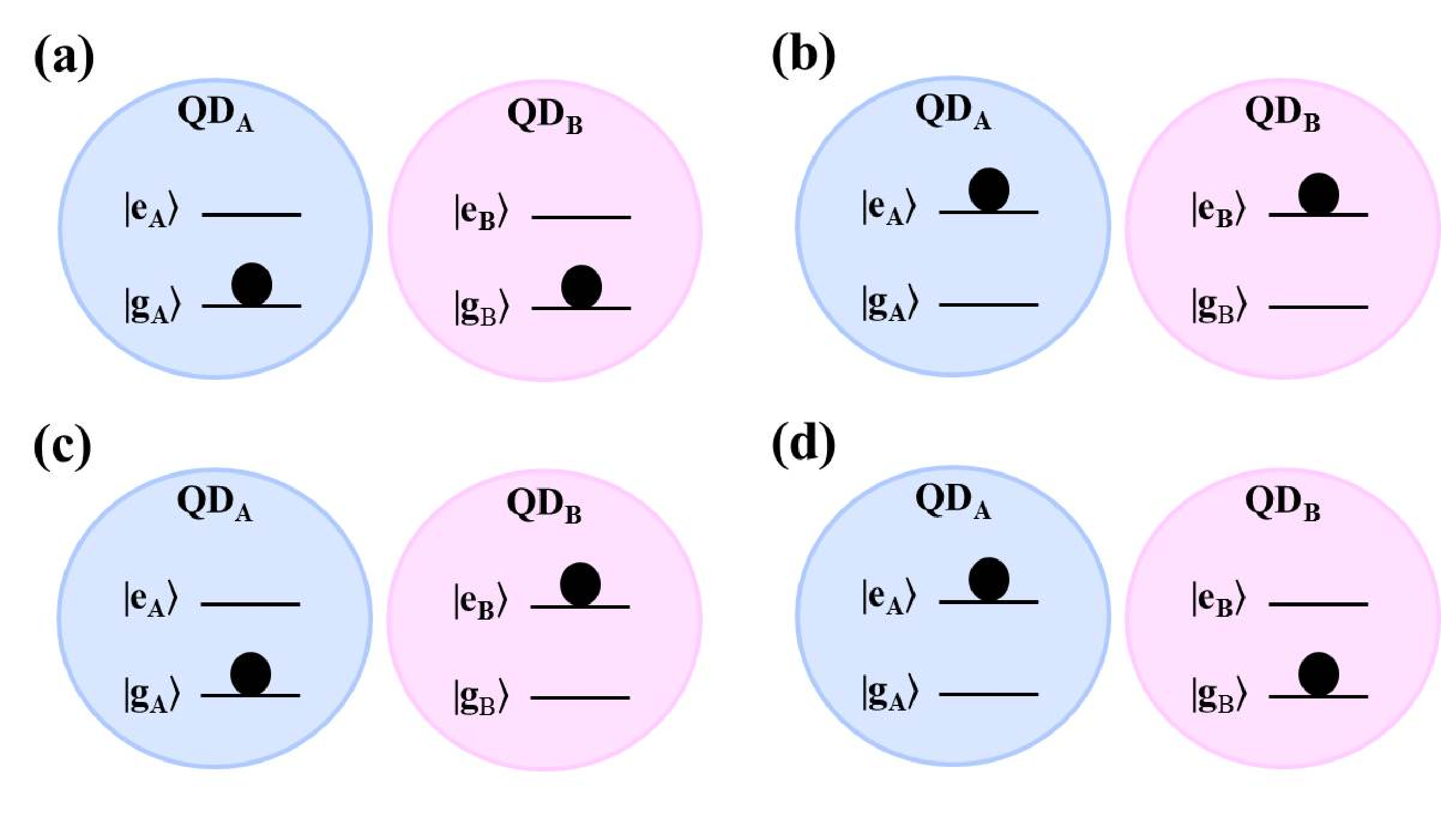}
\caption{Configuration of initial states of two spinless electrons of molecular double quantum dot}\label{Fig4-Configurations}
\end{figure}

Our proposed molecular system is occupied with total two spinless
electrons in four-dimensional single-particle Hilbert space. The
two-particle states of system can be written as
$|\psi\rangle_{AB}=|\Phi\rangle_A\otimes|\Phi\rangle_B$, in
which $|\Phi\rangle_A$($|\Phi\rangle_B$) shows the state of
qubit A(B). As each electron of each dot can capture either ground
or excited state, the general form of occupation states can be
represented as
$|g_A,e_A,g_B,e_B\rangle=|g_A,e_A\rangle\otimes|g_B,e_B\rangle$.
Fig.(\ref{Fig4-Configurations}) shows the configuration of all possible initial states of our two-spinless electron system. 

Here, we discuss about the influence of energy-dependent
coefficients on the entanglement of quantum dot molecular system. The energy
contributions which can be taken into account asymmetrically
originated from the QD-reservoir couplings. Indeed, the unequal left
and right superconducting energy gaps of reservoirs ($\Delta_L \neq
\Delta_R$) can intensively influence the behavior of entanglement.
According to Eq.(\ref{MasterEq}) and Eq.(\ref{dissipation-1}) the
energy-dependent of molecular system is affected by a set of
elements: distribution function, density of states and coupling
coefficients. To observe the prominent role of asymmetry on the
entanglement of QDM system, we consider the coupling coefficients and
superconducting energy gaps in right-left asymmetric situation.
For this purpose, the strength of coupling coefficients which
strongly depends on the properties of QDs can easily be tuned
left-right asymmetrically by mean of the relevant gap voltages.
Also, the superconducting energy gaps of left and right reservoirs
can be simply chosen unequally in the arrangement of setup. To
present the effect of asymmetric coupling coefficients, we define
the asymmetric factor as a function of coupling contributions:
 \begin{equation}\label{asymmetric factor}
\kappa=\frac{\kappa_{A}+\kappa_{B}}{2}
\end{equation}
in which $\kappa_{\alpha}=|\frac{T_{\alpha L}-T_{\alpha R}}{T_{\alpha L}+T_{\alpha R}}|$, $\alpha=A,B$. Here, $T_{AL}$ denotes the coupling of $QD_A$ to
the near-lead(Left Lead) and $T_{AR}$ shows the coupling of
this $QD$ to the far-lead(Right Lead). Similarly, the coupling of $QD_B$ with
the far-lead(Left Lead) is shown by $T_{BL}$ and with the near-lead(Right Lead) is indicated by
$T_{BR}$ which is  illustrated in Fig.(\ref{Fig1}). All these coupling parameters are considered positive which provide the magnitude of asymmetric factor from zero to unit.

Mostly, in the study of QDs system  for simplification, the coupling
of QD with the far-lead is ignored\cite{Segal-2015}. However, we
assume the both coupling of each QD to the near-lead and far-lead non-zero with only different strengths which are involved in the asymmetric
factor definition, (Eq.({\ref{asymmetric factor})).

According to the definition of asymmetric factor,
(Eq.(\ref{asymmetric factor})) and also refer to the configuration of the initial states(Fig.(\ref{Fig4-Configurations})), we investigate the entanglement of
our proposed quantum dot molecular system in two parts, namely symmetric and
asymmetric structures as follows.
\subsection{Symmetric Structure}
The initial state of symmetric structure is defined as the
superposition of states in $(c)$ or $(d)$ configurations in Fig.(\ref{Fig4-Configurations}) which can
provide Bell states. Particularly, we use the state
$\frac{1}{2}(|1,0,0,1\rangle-|0,1,1,0\rangle)$ for the symmetric
structure.

Also, in this structure, the left coupling coefficient of each QD is
similar to the right one ($T_{AL}=T_{AR}$ and $T_{BL}=T_{BR}$) which means the left-right symmetric coupling
coefficients. This situation supplies the minimum magnitude of
the asymmetric factor, $\kappa=0$. Also, this situation corresponds to
equal superconducting energy gaps of the left and right reservoirs ($\Delta_{L}=\Delta_{­R}$). In these conditions, the entanglement of QDM
system is obtained only for the initial entangled states.

For this purpose, we consider the following Bell state as an initial state with the
highest degree of entanglement:
\begin{equation}\label{Bell}
\rho(0)= \left[ {\begin{array}{cccc}
    0 & 0 & 0 & 0 \\
    0 & 0.5 & -0.5i & 0  \\
    0 & 0.5 i & 0.5 & 0 \\
    0 & 0 & 0 & 0 \\
  \end{array} } \right].
\end{equation}
\subsection{Asymmetric Structure}
The configuration of asymmetric structure can involves in one
of $(a)$ or $(b)$ in Fig.( \ref{Fig4-Configurations}) that
we choose $(a)$ configuration. This situation means that both
QDs are occupied with spinless electrons in ground states of each
QD.

Moreover, the asymmetric structure is defined for the left-right different
coupling coefficients with $0<\kappa\leq1$ magnitude and the unequal
order parameters of reservoirs, $\Delta_{L}\neq\Delta_{­R}$. In
this group, the ideal asymmetry element is achieved for the
maximum amount of asymmetric factor $\kappa \simeq 1$. The situation
of ideal asymmetry is available when one of the left or right
coupling coefficient is much larger than the other one. To apply the
ideal asymmetry properties in physically rational considerations, we
assume that each QD is coupled to the near-lead with much larger
strength than the far-lead. In other words, we consider
$\Gamma_{AL} \gg \Gamma_{AR}$ and $\Gamma_{BR} \gg \Gamma_{BL}$ to
realize the most magnitude of asymmetric factor. An interesting
feature in the composed systems is the realization of entanglement from
the initial unentangled states. This important point would be accomplished
in the asymmetric structure. To investigate this significant
situation in the present system, we assume an appropriate separated
initial state as:
\begin{equation}\label{initial-asymmetric}
\rho(0)= \left[ {\begin{array}{cccc}
   1 & 0 & 0 & 0 \\
    0 & 0 & 0 & 0  \\
    0 & 0 & 0 & 0 \\
    0 & 0 & 0 & 0 \\
  \end{array} } \right].
\end{equation}
In the next section, we present the concurrence behavior of the
present QDM system for both symmetric and asymmetric structures.
\section{Results}\label{Results}
In this section, we investigate the concurrence behavior of
molecular system firstly in response to bias voltage, secondly by
the time evolution in constant voltage and finally through the
dynamics for specific superconducting energy gaps.
\subsection{Concurrence-Voltage characteristics}
The concurrence-voltage (C-V) characteristics demonstrates the
response of concurrence to the bias voltage as an external
easy-tunable driving field. Fig.(\ref{Fig5}) shows C-V
characteristics for normal reservoirs with $\Delta=0$ and
superconducting ones with specific order parameters in the
conditions of symmetric structure (Panel(a)) and asymmetric one
(Panel(b)).
\begin{figure}[t]
\begin{minipage}[t]{0.8\linewidth}
    \centering
    \includegraphics[width=1\textwidth]{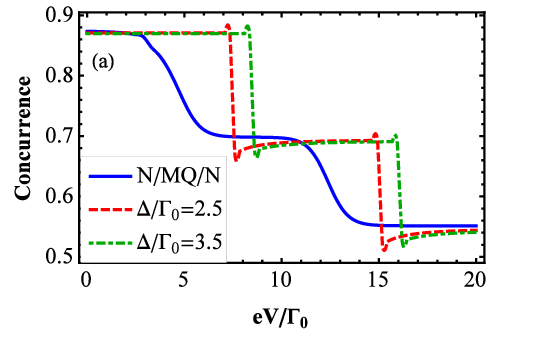}
   \end{minipage}
\hspace{0.1cm}
\begin{minipage}[t]{0.8\linewidth}
    \centering
    \includegraphics[width=1\textwidth]{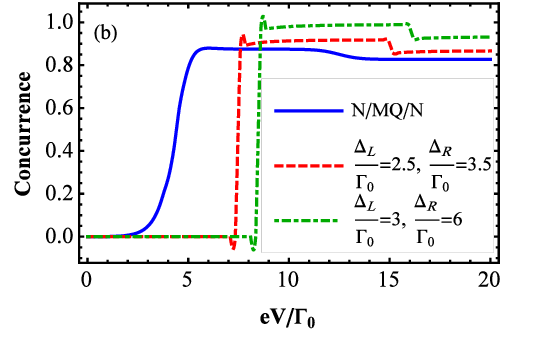}
      \end{minipage}
    \caption{The concurrence-voltage characteristics for the symmetric structure
 in panel (a) and for the asymmetric structure in panel (b).
 Normal leads:  Solid line $\Delta=0$, Superconducting leads: panel (a),
 Dashed line $ \frac{\Delta}{\Gamma_0} =2.5$, Dotted-dashed
    $\frac{\Delta}{\Gamma_0}=3.5$ ($\Delta_L=\Delta_R=\Delta$) and panel (b), Dashed
 line $\frac{\Delta_L}{\Gamma_0}=2.5$,
$\frac{\Delta_R}{\Gamma_0}=3.5$; Dotted-dashed
$\frac{\Delta_L}{\Gamma_0}=3$, $\frac{\Delta_R}{\Gamma_0}=6$. $\Gamma_0=\pi N_F |T|^2$ .}\label{Fig5}
\end{figure}
In panel (a) of Fig.(\ref{Fig5}), the concurrence shows degradation
for the symmetric structure while in panel (b) of this figure, the
concurrence indicates rising for the asymmetric conditions in higher 
values of bias voltage. In both panels of Fig.(\ref{Fig5}), the
concurrence changes in the energy levels of QDs with the step shapes
for the normal leads and with the delta peaks for the
superconducting reservoirs. The presence of superconductors as
reservoirs provides stronger response than the normal leads. This
effect is obvious in Fig.(\ref{Fig5}) when in panel (a) the
concurrence is decreased with higher values of potential and in
panel (b) concurrence shows increment in higher magnitude for the
Josephson junction than the normal one. Also, the influence of
superconducting reservoirs is displayed more clearly when by increasing the amount of
superconducting energy gaps, the concurrence has lower
magnitude for the symmetric structure (panel (a)) and inversely for
the asymmetric group (panel (b)).

It is interesting that our proposed setup is able to support two different fundamental concepts of
physics which are quantum transport and quantum entanglement
simultaneously in response to bias voltage changes. There are two
different responses to the increase of bias voltage in I-V
characteristics (Fig.(\ref{Fig3})) and C-V one
(Fig.(\ref{Fig5})-(a)) which are increasing for the former and
decreasing for the later in the symmetric structure. This various
behavior can be interpreted as the response of electrons to the
external bias voltage.
As a consequence of the bias voltage increasing, electrons are accelerated which cause a rising current across the symmetric junction.
This faster movement of carriers means that the electrons with the initial entangled states (Eq.(\ref{Bell})) in the symmetric coupling situation can be present in a shorter period of time which leads to have less time to be entangled. Therefore, the entanglement degradation in response to bias voltage rising does make sense for the left-right symmetric conditions.

The behavior of concurrence originates from the moving manner of electrons. So, it would be physically reasonable that carriers can move more quickly through the Josephson junction than the normal one which means the lower value of entanglement in JJ for the left-right symmetric situation.
However, the electrons of asymmetric structure with the initial unentangled states (Eq.(\ref{initial-asymmetric})) have different conditions.
The asymmetric coupling coefficients provide a bounded-like situation for the unentangled electrons which give them an opportunity to be well entangled. It means that although electrons can move faster by the increase of bias voltage, the left-right asymmetric situation arranges the possibility of being robustly entangled for them. Therefore, it would be logical that C-V characteristics shows increasing in response to bias voltage rising for asymmetric group.

It is a crucial point that although the concurrence of asymmetric structure(panel (b) of Fig.(\ref{Fig5})) behaves differently from the symmetric one(panel (a) of Fig.(\ref{Fig5})), their origins are the same.
The main reason for this variety behavior of concurrence refers to the powerful strength of our proposed setup in controlling the features to obtain the desired results.
In addition, the quantum correlation between the localized sites
(QDs) is under investigation for the present system. This
correlation is attributed to the presence of electron in each QD. In
a higher current magnitude, the electrons are passed faster through
the QDM system and consequently, the correlation between localized
sites is decreased.
\subsection{Dynamics of concurrence in constant bias voltage}
The time evolution of concurrence for given values of
superconducting energy gap in a constant bias voltage is
demonstrated for symmetric and asymmetric structures in panels
(a) and (b) of Fig.(\ref{Fig6}), respectively.
\begin{figure}[t]
\begin{minipage}[t]{0.8\linewidth}
    \centering
    \includegraphics[width=1\textwidth]{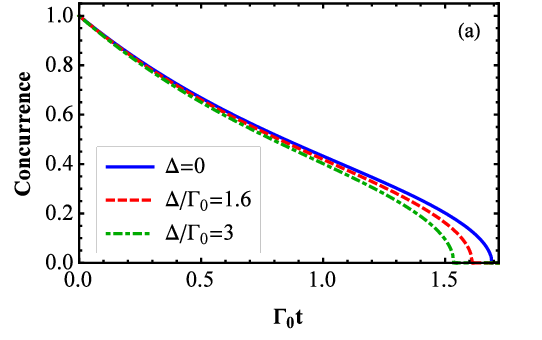}
\end{minipage}
\hspace{0.1cm}
\begin{minipage}[t]{0.8\linewidth}
    \centering
    \includegraphics[width=1\textwidth]{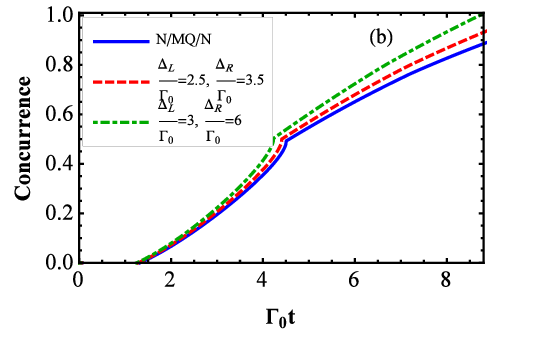}
\end{minipage}
 \caption{The dynamics of concurrence for the symmetric structure with constant low bias voltage in panel (a) and for the asymmetric structure with constant high bias voltage in panel (b).
 Normal leads:  Solid line $\Delta=0$, Superconducting leads: panel (a),
 Dashed line $ \frac{\Delta}{\Gamma_0} =1.6$, Dotted-dashed
 $\frac{\Delta}{\Gamma_0}=3$ ($\Delta_L=\Delta_R=\Delta$) and panel (b), Dashed
 line  $ \frac{\Delta_L}{\Gamma_0} =2.5$, $ \frac{\Delta_R}{\Gamma_0} =3.5$;
 Dotted-dashed $ \frac{\Delta_L}{\Gamma_0} =3$, $ \frac{\Delta_R}
 {\Gamma_0} =6$. $\Gamma_0=\pi N_F |T|^2$. }\label{Fig6}
\end{figure}
According to Eq.(\ref{Bell}) and Eq.(\ref{initial-asymmetric}) which
express the initial states of symmetric and asymmetric group
conditions, the concurrence of these structures are increased and
decreased through the time, respectively. In Fig.(\ref{Fig6}), the
concurrence of both left-right symmetric and asymmetric situations
decay faster for JJs than the normal ones to receive the ultimate
magnitudes only with opposite manner. Indeed, the decay rate of
concurrence is speeded up by increasing the superconducting energy
gap for them.
\subsection{Dynamics of concurrence for specific superconducting energy gaps}
Fig.(\ref{Fig7}) indicates the time evolution of entanglement for
energies which are in resonant with the energy levels of QDs,
$eV=\varepsilon_{i}+ \Delta\;(i=A,B)$, in symmetric structure
(panel (a)) and asymmetric one (panel (b)). These resonant points
are illustrated as peaks with respect to bias voltage in
Fig.(\ref{Fig5}). Due to the proximity effect of superconducting
reservoirs, the concurrence behaves differently in two sides of each
resonant points.
\begin{figure}[t]
\begin{minipage}[t]{0.8\linewidth}
    \centering
    \includegraphics[width=1\textwidth]{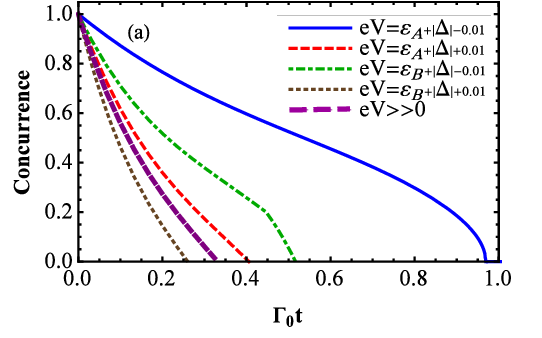}
\end{minipage}
\hspace{0.1cm}
\begin{minipage}[t]{0.8\linewidth}
    \centering
    \includegraphics[width=1\textwidth]{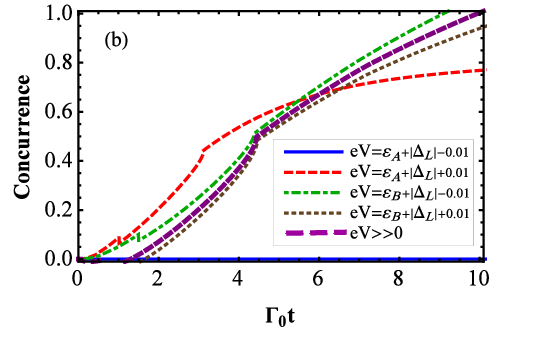}
\end{minipage}
 \caption{The dynamics of concurrence for bias voltages in resonant with QD's energy levels, panel (a): for the symmetric structure and panel (b) for the asymmetric structure. Solid line: left side of the first resonant point, Dashed line: right side of the first resonant point, Dot-dashed line: left side of the second resonant point, Dotted line: right side of the second resonant point and Thick-dashed line: high bias. $\Gamma_0=\pi N_F |T|^2$ (for panel (a) $\Delta_L=\Delta_R=\Delta$).}\label{Fig7}
\end{figure}
For the left-right symmetric group, concurrence shows longer elapsed
time for the left side of the resonant point ($eV=\varepsilon_{i}+
\Delta-0.01; i=A,B; \Delta_L=\Delta_R=\Delta$) than the right side
($eV=\varepsilon_{i}+ \Delta+0.01$) which is illustrated in panel
(a) of Fig.(\ref{Fig7}). However for high bias voltage $eV \gg 0$,
the dynamics of concurrence decays in moderate rate. For asymmetric
situation which is shown in panel (b) of Fig.(\ref{Fig7}),
the time evolution of concurrence for the first resonant point
($eV=\varepsilon_{A}+ \Delta_{L}$) which corresponds to the low bias
voltage shows different behavior than the second one
($eV=\varepsilon_{B}+ \Delta_{R}$).

The dynamics of concurrence for the left side of the first resonant
point ($eV=\varepsilon_{A}- \Delta_{L}-0.01 $) illustrates no response
through time and also it shows the longest decay rate for the right
side of this point($eV=\varepsilon_{A}+ \Delta_{L}+0.01 $). This
behavior means that for asymmetric situation, the amount of bias
voltage plays more important role than the other elements in the
concurrence time evolution. However for the second resonant point
with larger bias voltage, the dynamics of concurrence is similar to
the symmetric situation(panel (a) of Fig.(\ref{Fig7})) only with
inverse manner of decay. In other words, the dynamics of concurrence
shows longer elapsed time for the right side of the second resonant
point, $eV=\varepsilon_{B}+ \Delta_{R}+0.01$, than the left side,
$eV=\varepsilon_{B}+ \Delta_{R}-0.01$. Also for high bias voltage $eV
\gg 0$, it decays in moderately rate.
\section{Conclusion}\label{Conclusion}
In summary, we proposed a protocol to obtain perfect entanglement
for two coupled QDs molecular structure in a voltage-controlled
junction. In this strategy, we focused on the arrangement of
different controlling elements to enhance the quantum information
characteristics of system. First by engineering the reservoirs, we
applied superconductors as leads using the significant properties of
Josephson junction under the bias voltage control. Second, we
utilize the energy couplings of QD-reservoirs asymmetrically. The
main advantage of this hybrid quantum system refers to its wide
strength of controllability due to the easy tuning driven bias
voltage and also control of coupling coefficients by manipulating
the quantum dot barriers with respect to the required results.
In concurrence-voltage characteristics, applying
the asymmetric coupling energy conditions can provide high degree of
entanglement while for the symmetric situation, entanglement shows
degradation.

\section*{Appendix A: Diagonalizing the Hamiltonian of superconducting leads} \label{AppendixA}
It would be possible to diagonalize the superconducting Hamiltonian.
The mean-field Hamiltonian of superconducting leads is mostly diagonalized by Bogoliubov transformation. To this end, we consider
the following Bogoliubov transformation \cite{Ketterson}:
\begin{eqnarray} \label{BogoliubovT}
\hat{c}_{-k \nu \downarrow} &=& u_{k \nu} \hat{\gamma}_{-k \nu \downarrow} -v^{*}_{k \nu} \hat{\gamma}^{\dagger}_{k \nu \uparrow},
\nonumber \\
\hat{c}^{\dagger}_{k \nu \uparrow} &=& u^{*}_{k \nu} \hat{\gamma}^{\dagger}_{k \nu \uparrow} + v_{k \nu} \hat{\gamma}_{-k \nu \downarrow},
\end{eqnarray}
where $\hat{\gamma}^{\dagger}_{k \nu \sigma}(\hat{\gamma}_{k \nu
\sigma})$ denotes the creation (annihilation) operator of Bogoliubov
fermionic quasiparticle excitation. Bogoliubov quasiparticles follow
the fermionic anticommutation relation $ \{ \hat{\gamma}_{k \nu
\sigma}, \hat{\gamma}^{\dagger}_{ k^{'} \nu^{'} \sigma^{'} } \}
=\delta_{\nu \nu^{'} } \delta_{ k k^{'} } \delta_{\sigma \sigma^{'}
} $. The complex number parameters $u_{k \nu}$ and $v_{k \nu}$
adopting the relation $|u_{k \nu}|^2+|v_{k \nu}|^2=1$ are defined
as:
\begin{eqnarray}
u_{k \nu} &=&e^{-i \Phi_{\nu}} \sqrt{ \frac{1}{2} \left( 1+ \frac{\xi_{k \nu}}{|E_{\nu k}|} \right) },
\nonumber \\
v_{ k \nu} &=&\sqrt{\frac{1}{2} \left( 1- \frac{\xi_{\nu k}}{|E_{k \nu}|} \right)}.
\end{eqnarray}
Here, $E_{\nu k}=\sqrt{\xi^2_{k \nu}+|\Delta_{\nu}|^2}$ indicates
the quasiparticle energy. Inserting the Bogoliubov
transformation (\ref{BogoliubovT}) into the mean-field
Hamiltonian (\ref{MFH}), the diagonalized Hamiltonian is achieved:
\begin{equation}
\hat{H}_{Leads}=E_G+\sum_{k \nu \sigma} E_{\nu k}  \hat{\gamma}^{\dagger}_{k \nu \sigma} \hat{\gamma}_{k \nu \sigma},
\end{equation}
in which the ground state energy $E_G$ shows the Cooper pair condensate energy.
\section*{Appendix B: Dynamics of system} \label{AppendixB}
To study the dynamics of system, we start from the Liouville-von
Neumann equation of the complete system in the interaction picture
\cite{Breuer}. A comparison between the characteristics time scales
of the system, the QD relaxation time and the superconducting
coherence time as the environment time scale, implies that the
present system would be studied under the Markovian approximation
\cite{Grifoni,Baart,Harvey}. After partial tracing out the lead
degrees of freedom and applying the Born-Markov approximation, the
quantum master equation (QME) for the reduced density matrix is
obtained:
\begin{eqnarray}
\frac{d \hat{\rho}(t)}{dt}&=&-\frac{i}{\hbar}
[\hat{H}_{I},\hat{\rho}(t)]\\&-& \frac{1}{\hbar^2}\int^\infty_0
dt^{'} Tr_B \{ [\hat{H}_{I}(t),[\hat{H}_{I}(t^{'}),\hat{\rho} (t)]]
\},\nonumber
\end{eqnarray}
where $\hat{\rho}$ denotes the reduced density matrix of system in the interaction picture. The first term shows the Lamb shift
which is ignored in the present study and the second one represents
the dissipation of system.

In general case, the interaction Hamiltonian can be considered as
$\hat{H}_I=\sum_{\alpha} \hat{A}_{\alpha} \hat{B}_{\alpha}$ with the
operators $\hat{A}_{\alpha}$ and $\hat{B}_{\alpha}$ which satisfy the commutation relation $[A_{\alpha}, B_{\alpha}]=0$ and act on the system and leads Hilbert spaces, respectively. Finally,
the master equation for the central system of QDM in the presence of
superconducting leads is derived as:
\begin{eqnarray}\label{MasterEq}
\frac{d \hat{\rho}_s (t)}{dt}&=&\frac{1}{\hbar^2} \sum_{\omega}
\sum_{\alpha, \beta}\Big( \Gamma^{-}_{\alpha,\beta}(\omega) \Big(
\hat{A}_{\beta}(\omega) \hat{\rho}_s (t)
\hat{A}^{\dagger}_{\alpha}(\omega) \nonumber\\&-&\frac{1}{2} \{
\hat{A}^{\dagger}_{\alpha} (\omega) \hat{A}_{\beta}(\omega),
\hat{\rho}_s (t) \} \Big) \nonumber \\ &+&
\Gamma^{+}_{\alpha,\beta}(\omega) \Big(
\hat{A}^{\dagger}_{\alpha}(\omega) \hat{\rho}_s (t)
\hat{A}_{\beta}(\omega)\\ \nonumber &-&\frac{1}{2}
\{\hat{A}_{\beta}(\omega) \hat{A}^{\dagger}_{\alpha} (\omega),
\hat{\rho}_s (t) \} \Big )\Big),
\end{eqnarray}
where $\{ \}$ denotes the anticommutation relation. The dissipation
coefficients  $\Gamma^{-}_{\alpha,\beta}(\omega)=\int^{\infty}_0 ds
e^{i \omega s} \langle \hat{B}_{\alpha} (t)
\hat{B}^{\dagger}_{\beta}(t-s)\rangle_{B} $ and
$\Gamma^{+}_{\alpha,\beta}(\omega)=\int^{\infty}_0 ds e^{i \omega s}
\langle \hat{B}^{\dagger}_{\beta}(t)
\hat{B}_{\alpha}(t-s)\rangle_{B}$ are related to the bath
correlation function. For superconducting leads, the distribution
function is defined as $\langle \hat{\gamma}^{\dagger}_{k,\nu}
\hat{\gamma}_{k,\nu}\rangle_B =\frac{1}{e^{\beta
E_{k,\nu}}+1}=f^{+}(E_{k,\nu})$ and also $\langle
\hat{\gamma}_{k,\nu}
\hat{\gamma}^{\dagger}_{k,\nu}\rangle_B=(1-f^{+}(E_{k,\nu}))=f^{-}(E_{k,\nu})
$. So, we have
\begin{eqnarray}\label{dissipation-1}
\Gamma^{+}_{\alpha,\beta}(\omega)&=& 2 \pi \sum_{k\nu\sigma} T_{k\nu\alpha} T^{*}_{k^{'}\nu^{'}\beta} \langle  \hat{\gamma}^{\dagger}_{k\nu\sigma}
\hat{\gamma}_{k^{'}\nu^{'}\sigma^{'}} \rangle\nonumber \\&=& 2 \pi |T_{k\nu\alpha}|^2 \int dE f^{+}(E_{k,\nu}) R_{k\nu}(E_{k\nu}),
\nonumber \\
\Gamma^{-}_{\alpha,\beta}(\omega)&=& 2 \pi \sum_{k\nu\sigma} T_{k\nu\alpha} T^{*}_{k^{'}\nu^{'}\beta} \langle \hat{\gamma}_{k\nu\sigma}
\hat{\gamma}^{\dagger}_{k^{'}\nu^{'}\sigma^{'}}\rangle_B\nonumber\\&=& 2 \pi
|T_{k\nu\alpha}|^2  \int dE f^{-}(E_{k,\nu}) R_{k \nu}(E_{k\nu}).
\end{eqnarray}
According to the BCS theory, the superconducting density of states
$R_{\nu}(E)$ is defined \cite{Kosov,Grifoni,Afsaneh}:
\begin{equation}
R_{k \nu}(E)=N_F \frac{|E_{k \nu}|}{\sqrt{E_{k \nu}^2 - |\Delta_{\nu}|^2}},
\end{equation}
in which $N_F$ denotes the density of states for normal reservoirs
which is assumed as a constant parameter close to the Fermi level of
energy.
We define $\Gamma_0=2 \pi N_F |T_{k\nu\alpha}|^2$ so,
Eq.(\ref{dissipation-1}) can be written as:
\begin{eqnarray}\label{dissipation-2}
\Gamma^{+}_{\alpha,\beta}(\omega)&=&\Gamma_0 \int dE
f^{+}(E_{k,\nu}) \frac{|E_{k \nu}|}{\sqrt{E_{k \nu}^2 -
|\Delta_{\nu}|^2}}, \nonumber \\
\Gamma^{-}_{\alpha,\beta}(\omega)&=&\Gamma_0 \int dE
f^{-}(E_{k,\nu}) \frac{|E_{k \nu}|}{\sqrt{E_{k \nu}^2 -
|\Delta_{\nu}|^2}}.
\end{eqnarray}
To parameterize the effect of left-right asymmetric coefficients in Eq.(\ref{dissipation-2}), we define $T_{k \nu \alpha}=\gamma_{\alpha,\nu}T_0$ in which $\gamma_{\alpha,\nu}$($\nu=L,R $, $\alpha=A,B$) denotes the asymmetric constant parameter and $T_0$ shows the symmetric coupling coefficient.

In Eq.(\ref{MasterEq}) $\hat{A}_{\alpha}(\omega)$ denotes the
projection super-operator which acts on the eigenoperator of system with eigenvalue of $\omega$. Here, as we encounter with a
bipartite central system, we introduce the eigenoperator as $ |
\omega \rangle =| \varepsilon_A, \varepsilon_B
\rangle=|\varepsilon_A \rangle_A \otimes |\varepsilon_B \rangle_B $
with eigenvalue $ \omega= \{ \omega_A, \omega_B \}$. Therefore, we
define the super-operator:
\begin{equation}
\hat{A}(\omega)=\hat{A}(\omega_i, \omega_j)=
\sum_{
\begin{array}{c}
 \omega_i=\varepsilon^{'} _{i} - \varepsilon_{i}  \\
 \omega_j=\varepsilon^{'} _{j} - \varepsilon_{j}  \\
  \end{array}
} |\varepsilon_i \varepsilon_j \rangle \langle \varepsilon_i
\varepsilon_j | \hat{A} | \varepsilon^{'} _{i} \varepsilon^{'} _{j}
\rangle \langle \varepsilon^{'} _{i} \varepsilon^{'} _{j} |.
\end{equation}
The computational basis which is applied for the present system includes $ |1\rangle = |g_A,g_B\rangle$, $ |2\rangle=| g_A,e_B\rangle$, $|3\rangle=|e_A,g_B\rangle$ and $|4\rangle =| e_A,e_B\rangle$ where $|g_{\alpha} \rangle $ and $|e_{\alpha}\rangle $ represent the ground and excited states of quantum dots respectively ($\alpha=A,B$).

To consider the weak-coupling regime, the energies of system should be under the relation of $\Gamma^{\pm}_{\alpha,\beta} < \varepsilon_A, \varepsilon_B, |\Delta|$.\\ 
\section*{References}

\end{document}